# Using the C4ISR Architecture Framework as a Tool to Facilitate VV&A for Simulation Systems within the Military Application Domain

*Andreas Tolk, Ph.D.*

Virginia Modeling Analysis and Simulation Center
College of Engineering and Technology
Old Dominion University
7000 College Drive
Suffolk, Virginia 23435
Voice (757) 686 – 6203
Fax (757) 686 – 6214
Email: atolk@odu.edu

*Abstract*

*To harmonize the individual architectures of the different commands, services, and agencies dealing with the development and procurement of Command, Control, Communications, Computing, Surveillance, Reconnaissance, and Intelligence (C4ISR) systems, the C4ISR Architecture Framework was developed based on existing and matured modeling techniques and methods. Within a short period, NATO adapted this method family as the NATO Consultation, Command, and Control (C3) System Architecture Framework to harmonize the efforts of the different nations. Based on these products, for every system to be fielded to be used in the US Armed Forces, a C4I Support Plan (C4ISP) has to be developed enabling the integration of the special system into the integrated C4I Architecture.*

*The tool set proposed by these architecture frameworks connects operational views of the military user, system views of the developers, and the technical views for standards and integration methods needed to make the network centric system of systems work. The tools are therefore logically a valuable backbone for Verification, Validation, and Accreditation (VV&A). Their application is not limited to C4ISR systems; they can be used to define requirements and connected solutions and algorithms of Modeling and Simulation (M&S) systems as well. Especially for M&S systems to be used in connection with C4ISR system, the use of the C4ISR Architecture Framework would not only be a help, but can nearly be seen to be necessary to avoid double work and foster reuse and interoperability from the first stages of a project on. To enable the reader to build his own picture, the respective tools used and their application in the context of VV&A will be explained in form of an overview.*

## 1   Introduction to the C4ISR Architecture Framework

This paper is based on the follow-on of the work presented during the Simulation Interoperability Workshop Fall 2001 [Adshead et al., 2001]. Especially this first section reuses a lot of the explanations already given in section 3 of this reference.

Until recently, the individual commands, services, and agencies developed their C4ISR architectures using techniques, vocabularies, and presentation schemes that suited their unique needs and purposes. With the increasing focus on joint and combined operations on the one side, as well as the government-wide streamlining and downsizing initiatives on the other side, opportunities for cross-organization and collaboration arose. Architectures provide the interoperability framework, and the next logical step was to look for a common C4ISR Architecture Framework.

In October 1995, the Deputy Secretary of Defense directed a DoD-wide effort to define and develop better means and processes to ensure that C4ISR capabilities meet the needs of the warfighters. The establishment of the C4ISR Integration Task Force (ITF) under the direction of the Assistant Secretary of Defense for Command, Control, Communications, and Intelligence (ASD/C3I) consecutively let to the first version of a common C4ISR Architecture Framework in June 1996.

In October 1996, the C4ISR Architecture Working Group (AWG) was established to continue the work. New products and tools already being in productive use in the four different services enriched the first version. In addition to the four Services and Command representation, participants included

- **BMDO:** Ballistic Missile Defense Organization;
- **CISA:** C4I Integration Support Activity;
- **DARO:** Defense Airborne Reconnaissance Office;
- **DIA:** Defense Intelligence Agency;
- **DISA:** Defense Information Systems Agency;
- **DMSO:** Defense Modeling and Simulation Office;
- **JBC:** Joint Battle Center;
- **JTAMDO:** Joint Theater Air and Missile Defense Organization;
- **NIMA:** National Imagery Management Agency;
- **OASD (C3I):** Office of the Assistant Secretary of Defense for C3I;
- **OUSD (A&T):** Office of the Undersecretary of Defense for Acquisition and Technology;
- **SIMO:** Systems Integration Management Office

The leveraging of the prior efforts by those participants resulted in Version 2.0 released in December 1997. It is worth mentioning explicitly that beside all C4ISR related institutions already in 1996 the Defense Modeling and Simulation Office (DMSO) was included to insure the appropriate integration of Modeling and Simulation issues – something still not being taken fully into account in the international community (see, e.g., [Tolk, 2001]). This was done to ensured the applicability of the products not only to C4ISR systems, but also to simulation systems of the appropriate fidelity.

The Technical Working Groups of the North Atlantic Treaty Organizations soon realized the potential of this work. Therefore, they started to incorporate these ideas into the NATO Consultation, Command and Control Technical Architecture (NC3TA). The NC3TA describes the information technology (IT) basis to be used in NC3 systems in terms of architectural descriptions, reference models, and Off-The-Shelf (OTS)-technologies. Furthermore, the NC3TA integrates the technical aspects of specific architectures and frameworks developed by NATO such as the "NATO Information Security (InfoSec) Framework" and the "Architecture Description Framework" of the "NATO Interoperability Management Plan" (NIMP). It should therefore contribute to a better co-ordination within the NC3 Organization (NC3O). The NC3TA consists of five volumes:

- Management,
- Architectural Models and Description,
- Base Standards and Profiles,
- NATO C3 Common Standards Profile (CSP),
- NATO C3 Common Operating Environment (COE).

In November 2000, the US DOD C4ISR Architecture Framework [US DOD, 1997] was adapted as the NATO C3 System Architecture Framework [NATO, 2000] in a slightly modified form. Beside a different order and numbering of the respective products being used to describe the views, there are no main differences. In the appendix of [Adshead et al., 2001], a mapping matrix is given for the most often used products of the US version versus the NATO version. Within the applications known by the author, no substantial mismatches ever occurred so far, i.e., architectures having been delivered using the US DOD C4ISR Architecture Framework could easily be converted into a NATO version using the NATO C3 System Architecture Framework. Although speaking about the US DOD C4ISR Architecture Framework in the rest of the document, the results can be mapped directly to NATO applications as well.

## 1.1 Overview

The main objective of a common C4ISR architecture framework is to define a coordinated approach for C4ISR architecture development, representation, and integration by taking into account the evolutionary nature of framework development. The resulting C4ISR Architecture Framework is intended to ensure that the architectures developed by the geographic and functional unified commands, military services, and defense agencies are inter-relatable between and among the organizations' operational, systems, and technical architecture views, and are comparable and are capable of being integrated across joint and combined organizational boundaries. The C4ISR Architecture Framework has to be seen in the light that the government legislation is placing more emphasis on the need to pursue interoperable, integrated, and cost-effective business practices and capabilities within each organization and across the domains of defense. It provides direction on how to describe architectures by providing a "product-focused" method for standardizing architecture descriptions. The products are directly related to three different architecture views. The following definitions (in italic) can be found in [US DOD, 1997]:

- *The **operational architecture view** is a description of the tasks and activities, operational elements, and information flows required to accomplish or support a military operation.* In other words, the products describing the operational view catch the user requirements. Within this view, it is described WHAT has to be done in an operational context, i.e., what functionality is needed by the warfighter!

- *The **systems architecture view** is a description, including graphics, of systems and interconnections providing for, or supporting, warfighting functions.* The products being used in the system view describe the functionality of the systems – which already can be in place or under development – in stand-alone operations as well as in the integrated context. In other words, the HOW is defined in form of available functionality provided by the network centric system of systems.

- *The **technical architecture view** is the minimal set of rules governing the arrangement, interaction, and interdependence of system parts or elements, whose purpose is to ensure that a conformant system satisfies a specified set of requirements.* The products of the technical view are the home for standards to be used to enable the system of systems delivering the functionality described in the systems view.

In order to reach the high objective to provide mechanisms for incorporating information technology consistently, controlling the configuration of technical components, and ensuring compliance with the technical building codes within a joint and combined domain, all three architecture views have to be specified in sufficient detail. However, the level of detail can be tailored due to the fidelity of the models and systems being used. Therefore, e.g., no technical details concerning the system components have to be defined at the early stage of a project, but respective views have to be added later. This is consistent with the idea of evolutionary system development. The application of the C4ISR Architecture Framework ensures the appropriate level of detail in all participating views.

Within the next section, the products are enumerated. However, the C4ISR Architecture Framework comprises approximately 100 Pages describing 22 framework products plus some additional information on universal reference products (not counting the appendixes). Therefore, only a small idea of the potential of the C4ISR Architecture Framework can be given and it is strongly recommend to review the respective documents.

To summarize, the C4ISR Architecture Framework enables operationally driven solutions increasing the efficiency of the warfighter in user driven domains. By introducing various views, the user doesn't have to specify technical details. He can specify his requirements in a form easy to read and to understand by the users as well as by the system developers.

## 1.2 C4ISR Architecture Products

The C4ISR Architecture Framework [US DOD, 1997] specifies essential as well as supporting framework products and gives additional universal reference sources to be taken into account respectively. Four categories are distinguished:

- Products supporting the operational architecture view are numbered as OV-# products.
- Products supporting the systems architecture view are numbered as SV-# products.
- Products supporting the technical architecture view are numbered as TV-# products.
- Products giving a holistic view of the architecture (high level descriptions) are numbered as AV-# products.

Following essential products are described in [US DOD, 1997]:

| | |
|---|---|
| (AV-1) | Overview and Summary Information |
| (AV-2) | Integrated Dictionary |
| (OV-1) | High-Level Operational Concept Graphic |
| (OV-2) | Operational Node Connectivity Description |
| (OV-3) | Operational Information Exchange Matrix |
| (SV-1) | System Interface Description |
| (TV-1) | Technical Architecture Profile |

Following supporting products are described in [US DOD, 1997]:

| | |
|---|---|
| (OV-4) | Command Relationship Chart |
| (OV-5) | Activity Model |
| (OV-6) | Operational Activity Sequence and Timing Descriptions |
| (OV-7) | Logical Data Model |
| (SV-2) | Systems Communications Description |
| (SV-3) | System$^2$ Matrix |
| (SV-4) | System Functionality Description |
| (SV-5) | Operational Activity to System Function Traceability Matrix |
| (SV-6) | System Information Exchange Matrix |
| (SV-7) | System Performance Parameters Matrix |
| (SV-8) | System Evolution Description |
| (SV-9) | System Technology Forecast |
| (SV-10) | Systems Activity Sequence and Timing Descriptions |
| (SV-11) | Physical Data Model |
| (TV-2) | Standards Technology Forecast |

With only a few exceptions, respective products can be found in the NATO version also.

As universal reference resources, the following products are mentioned, but the level of detail in the description in [US DOD, 1997] is not sufficient to apply them. However, the reference list in [US DOD, 1997] helps to find the real documents needed in case the evaluation of the short descriptions give a hint that the respective product may be useful: the C4ISR Core Architecture Data Model (CADM), the Defense Data Dictionary System (DDDS), the Levels of Information Systems Interoperability Model (LISI), the Universal Joint Task List (UJTL), the Joint Operational Architecture (JOA), the DoD Technical Reference Model (TRM), the Defense Information Infrastructure Common Operating Environment (DII COE), the Shared Data Environment (SHADE) and the Joint Technical Architecture (JTA). It should be mentioned, that the C4ISR Core Architecture Data Model (CADM) is a data model enabling the storage and management of architecture definitions. It is not a data model to be used within a C4ISR system, but a data model describing the description of the overall architecture. (The data models used within the respective systems are themselves entities within the CADM).

It is interesting that - although DMSO has been involved in the process of bringing together the actual version of the C4ISR Architecture Framework - the aspect of Technical Reference Models of the Modeling and Simulation community or the aspect of simulation standards like Distributed Interactive Simulation (DIS) [IEEE 1278] or the High Level Architecture (HLA) [IEEE 1516] are not mentioned explicitly in the list of universal reference resources. Once again, as already shown in [Tolk, 2001], the effect that the standards of the M&S community are not taken into account strongly enough - from the beginning as well as in the operational frameworks and in Technical Reference Models - becomes obvious. That is why later they are unlikely to be used by the M&S community. The same is true the other way round as well, i.e., the C4ISR standards are not integrated into M&S solutions also.

Anyhow, [Adshead et al., 2001] shows that the C4ISR Architecture Framework was successfully applied to define C4ISR Architectures for NATO's operational concept of Active Layered Theater Ballistic Missile Defense (ALTBMD) by using matured M&S systems like the Extensive Air Defense Testbed (EADTB), Extensive Air Defense Simulation (EADSIM), and the Theater Missile Defense Simulator (TMDSIM) for the evaluation. It was used to connect the operationally driven user requirements formulated in the NATO Staff Target for ALTBMD with implementing systems (like PATRIOT or MEADS) and interoperability standards (like LINK 16) enabling an efficient discussion between all communities interested in the study and the results.

### 1.3  Applying the C4ISR Architecture Framework

Before showing the tight connection of a common architecture framework and VV&A, the application of the framework should be understood. The C4ISR Architecture Framework defines the following entities that are used within its products:

- An entity within the operational architecture view is an **operational node**. This operational node performs activities. An **activity** needs time and has to be sequenced and harmonized with other activities. This is a description of the operational needed functionality of the overall system without specifying technical details.
  E.g., instead of requiring a response time of the fire control system for a missile to be under 30 seconds, the requirement is formulated that the fire control system must react fast enough to engage an incoming missile at least twice. The 30 seconds timeframe should result from respective calculations that can be adjusted in case of technological changes, e.g., faster incoming TBM are the introduction of new technologies, like an Air-Defense-Laser.

- An entity within the systems architecture view is a **system**. A system performs functions. A **function** needs time and has to be sequenced and harmonized with other functions. Altogether, this leads to a description of the available functionality when all systems are in place and interconnected based on the standards defined in the technical view.

- The technical component of the C4ISR architecture is the set of rules that governs system implementation and operation, i.e., most often the identification of applicable portions of existing technical guidance, e.g., which exact protocol has to be used for a given information exchange over a communication link.

The operational nodes and their tasks and operational activities can be directly derived from the operational requirements. The warfighter defines the command hierarchy, the functionality needed, constraints for harmonization, connectivity of subsystems, etc.

The system developer – who can be an M&S developer, as the functionality can be delivered by a simulation system in the context of Simulation Based Acquisition (SBA), Training and Education, as well as to Support Real Military Operations – uses the system view products to describe the functionality of the system as well as the products of the technical view to assure the needed connectivity and timely passing of information.

Of special importance is the **Operational Activity to System Function Traceability Matrix**, the product SV-5 in the C4ISR Architecture Framework. This matrix shows, which system function delivers which operational activity, in other words, how does an actual architecture fulfill the operationally driven user requirements.

To summarize, a C4ISR architecture being described consistently and completely using the products of the C4ISR Architecture Framework is understandable for of participants. It can easily be shown, which systems function are necessary to meet the user requirements. It is also possible to show which new capabilities are achievable due to improvement of systems or the introduction of new technologies. Due to the network centric approach, synergy effects can be derived as well as unintended interrelations, e.g., in form of electromagnetic malfunctions due to improper system mixes used within a scenario.

## 2   The C4ISR as a Tool for VV&A in the Military Domain

This article wants to introduce the C4ISR Architecture Framework as a valuable tool to enable effective and efficient VV&A for M&S systems within the military application domain. To this end, in this section some challenges will be dealt with that VV&A has to cope with. It would go beyond the scope of this article to introduce VV&A; therefore, the interested reader is referred to general introductions like, e.g., available on the websites of the Defense Modeling and Simulation Office [DMSO, 2002]. A vision of the future of VV&A and what has to be taken into account is given in [Balci et al., 2002]. Many of the following subsections – in particular in the second half (starting with subsection 2.6) – are issues of the vision described in Balci's article.

However, it should be pointed out that this section is far away from being complete. It is just meant to show the tremendous potential lying within the use of a common architecture framework like the C4ISR Architecture Framework for VV&A purposes. Furthermore, the insights given in the next subsections are not based on broad experience but on the limited insights and ideas derived from the application of the C4ISR Architecture Framework in an M&S intensive context as described in [Adshead et al., 2001]. Additional R&D is therefore recommended.

### 2.1   Catching user requirement appropriately

The methods and products of the operational architecture view are easy to understand and easy to apply and enable to cope with user requirements of the Warfighter in an effective and efficient manner. Interrelations and additional assumption, hidden constraints and unintended influences become obvious and can be discussed on the basis of a common model between user and developer. The importance of system focused decisions in the overall context can be made clear to the decision makers on all levels. VV&A engineers can assure that the decisions and assumptions are caught and documented using the products of the C4ISR Architecture Framework enabling all participants to understand the requirements and how they are met by the proposed – and later realized – system of systems.

### 2.2   Tracing user requirement transparently

As the operational requirements are directly connected to the operational nodes and activities of the operational architecture view of the C4ISR Architecture Framework products, which are connected to the systems and their functions of the system architecture view, and the interconnectivity of the systems and components based on the standards described in the technical architecture view, every system function and every interconnection standard can be related to respective driving requirements. Functions not driven by a requirement are not necessary and give room for saving effort and time. Interconnections necessary but not supported by standards show the necessity for extra VV&A efforts as well as future R&D efforts to close the gap. That this is possible on a grater scale has been showed in [Adshead et al., 2001].

Another aspect is interesting in the light of the ongoing transformation of the US forces from threat based to capability-based evaluations. It doesn't matter whether the requirements are derived from an existing threat or are derived to fulfill future needs in the sense of improving the overall capabilities of forces. In addition to this traditional "top down" approach – leading from an operational concept to needed systems to make the vision come true –, the C4ISR Architecture Framework allows to build up new operational concepts in a "bottom up" manner as well, as it is possible to start with available systems and derive a respective operational architecture view. VV&A can assure that all relations and interconnections are derived properly and the operational and system architecture views are derived in the correct manner.

## 2.3 Validation of the derived models

As defined in the US DOD Instruction 5000.61, validation is the *process of determining the degree to which a model (and data) is an accurate representation of the real world from the perspective of the model's usage*, i.e., answering the question, if the right thing is modeled. As the various models are accessible and assessable in form of standardized models using standardized transformation processes, the validation in case of using the C4ISR Architecture Framework is indeed easier than in the usual case, i.e., the problem entity, the derived conceptual model, and the computerized model are using individual and not related forms in which they are captured.

## 2.4 Verification of the derived models

Again, following the US DOD Instruction 5000.61, verification is the *process of determining that a model implementation accurately represents the developer's conceptual description and specifications*, i.e., answering the questions, if the thing is modeled (in the sense of coded) correctly. Again, the use of common methods and products on all levels facilitate the application of usual verification tools.

## 2.5 Support of V&V techniques

The C4ISR Architecture Framework neither mandates nor forbids the application of any V&V techniques, like, e.g., explained in [Balci, 1998]. The presence of common models on the various levels of modeling with well-defined transformations, however, facilitates especially the application of informal, static, and formal techniques.

## 2.6 Expanding VV&A from Product-Centered Assessment to Product/Process/Project-Centered Assessment

As stated in [Balci, 2002], much of the M&S VV&A research and practice have concentrated on product measurement and evaluation. The same has been true for system procurement in the domain of C4ISR until recently also. With the introduction of the C4ISR Architecture Framework it is now possible, to evaluate the performance of a system in a much broader context. As relations and different roles of a system in the overarching architecture are captured by standardized means, the evaluation of the contribution of a new system – or the improvement of the functionality of a legacy system – to the target architecture is feasible. This framework can be easily used for VV&A purposes as well. Respective V&V procedures can be applied to ensure the correctness of assumptions and constraints and document them for later improvement respectively.

## 2.7 Developing M&S Applications Using the Component-Based Technology

Composability is an issue for C4ISR as well as for M&S systems. Actually, both communities are still working on the basics, although in the commercial information technology respective support already is an issue for a long time. The importance for M&S VV&A is documented in [Balci, 2002]. The C4ISR Architecture Framework can contribute to the discussion by introducing a standard way to model the components in all aspects relevant for composability. A component described by the means of the C4ISR Architecture Framework can be (re)used in the C4ISR domain as well as in the M&S domain. Interfaces, behavior, temporal constraints, logical interdependencies, relations to other elements, hidden constraints and assumptions can be modeled and made transparent to the user as well as the system developer. Many products are additionally close to respective views as proposed by software engineering standards, especially the Unified Modeling Language (UML).

### 2.8  Enabling Computer-Aided Support for VV&A

The neighborhood of many C4ISR Architecture Framework products to UML views enables computer-aided support in the process of creating and evaluating the respective views. Actually, the products being used in the NATO feasibility study underlying the insights described in [Adshead et al., 2001] are massively based on UML diagrams. This support can be added to the over 200 utilities and tools available in the commercial market providing computer-aided support for software testing mentioned in [Balci, 2002]. The C4ISR domain is only a small fraction of the market for information technology, but the C4ISR share is still considerably larger than the M&S fraction. Therefore, it only can be recommended to benefit from these efforts instead of looking for special M&S tools.

### 2.9  Measurement of the efficiency of VV&A

The value of VV&A and how to measure it is an issue of many practice-oriented articles. The C4ISR Architecture Framework doesn't answer questions like "How much VV&A is needed?" or "How much VV&A is enough?" However, the products of the C4ISR Architecture Framework show, what systems will be effected if one part of the architecture fails, what requirements cannot be met any longer if some functionality is not available any longer, and what functions cannot be provided to the other systems if some special standards necessary to insure the interconnectivity fail. To this end, at least a sort of risk management becomes inherently possible.

### 2.10  Effective Techniques for Utilizing Subject Matter Expert Knowledge

As stated in [Balci, 2002], we are facing major challenges in eliciting, representing, and integrating the knowledge of subject matter experts (SME) employed for VV&A tasks. Utilizing the products of the C4ISR Architecture Framework, the gap between the military user, the system developer, and the SME becomes a little bit smaller, as they can use common and easy to understand models to discuss their different points of view.

In addition it should be mentioned, that it maybe useful to add additional views to the C4ISR Architecture Framework. E.g., in [Adshead et al., 2001] the use of additional "Evaluation Views" for M&S purposes is mentioned. It is possible, that the SME Knowledge can be captured and encapsulated into products of a V&V architectural view. Such a methodological approach would not only help the M&S community, but would be a great benefit to the C4ISR community as well.

### 2.11  Contributing to a Common Terminology/Ontology

The necessity to create a common understanding is the topic of legions of interoperability related papers. In the context of VV&A it is necessary, to deal with the static aspects as well as the dynamic aspects of a system. Therefore, not only a common dictionary or data model is necessary, but the temporal and causal interdependencies of the various systems have to be understood as well. Actually, the Functional Description of the Mission Space – which complements the High Level Architecture and the Data Standards of DMSO to a Common Technical Framework (CTF) – tries to resolve this issue for the M&S community. Using the products proposed by the C4ISR Architecture Framework would not only help to make the results usable to the C4ISR community, also additional sources of knowledge would become usable for DMSO and the M&S community as well. Literally, every fielded system specified using the C4ISR Architecture Framework would immediately be available to the M&S community as well.

### 2.12  Improving the relationship between the engaged community

Last but not least, the VV&A team would improve the C4ISR Architecture Framework product. The VV&A experts would no longer be perceived as a sort of controller team of the managers, but they are part of the process to create a valuable product. The necessary independence of the VV&A agent isn't afflicted by this, as he still evaluates the product concerning the meeting of all requirements, but as he is working within the system – using the same framework and talking the same language –, his findings and recommendations are much more easily transferable to the rest of the team. The VV&A agent becomes a VV&A engineer who is perceivable for everybody contributing to the process of creating a valuable product, improving instead of criticizing and enabling instead of evaluating. As his work is measurable in form of additional functionality (or assure functionality), it is perceived as being of greater value as it is the case today, shifting the question "How much VV&A is necessary?" – often perceived as the necessary evil – to "How do we efficiently assure the availability of operationally necessary functionality of the system?"

## 3   Operational Aspects of a Common C4ISR Architecture

To close with, an additional operational aspect has to be taken into account as well. The procurement process for military systems is described in the US DOD 5000 Model series. Every new system the developers want to field has to be described using the products of the Command, Control, Communications, Computing, and Intelligence (C4I) Support Plan (C4ISP) first. The required products are a subset of the products defined in the C4ISR Architecture Framework. The C4ISP guidance is therefore introduced as follows [US DOD, 2002]:

*"The trend in modern warfare is toward the increased use of smart weapons and the integration of Command, Control, Computers, and Communications (C4) systems with Intelligence, Surveillance, and Reconnaissance (ISR) systems to maximize combat effectiveness. The complexity and cost of these integrated combat support systems are such that judgments regarding their design and procurement should be supported by the end-to-end analysis of the data/intelligence and infrastructure required to employ these new systems. Moreover, the results of this analysis should support the acquisition process; a specific goal is to incorporate C4I infrastructure early on into the acquisition design space, particularly with regard to overall system efficiency and supportability (i.e., consider the C4I infrastructure during design tradeoffs, in a manner similar to the consideration given the logistics infrastructure). Accordingly, the C4I Support Plan (C4ISP) evolved as a tool to identify, plan, and manage implementation issues related to C4I infrastructure for a specific system / program considering its entire acquisition and operational life cycle. The plan should, of necessity, be developed using a methodology, which will ensure that both the C4I infrastructure enabling the system, and the data provided to the system, are derived from specific operational requirements. Overall, the planning process should provide a thoughtful approach toward defining requirements, identifying shortfalls, and proposing solutions and their costs."*

Especially when M&S components are used to deliver functionality to support military operations, like Alternative Courses of Action Analyses, they become a part of the overarching C4ISR world. Also when being used for training and education, the M&S systems have to be coupled with the C4ISR systems to follow the "Train as you fight" philosophy, hence becoming partners in the C4ISR world. Last but not least, to objectives and purposes having lead to the C4ISP Guidance for C4ISR systems can be mapped easily to M&S systems as well. Efforts like composability are conducted in both worlds, reuse is an issue for C4ISR as well as M&S, and interoperability is a general challenge for both sides.

Moreover, there are some important benefits for the military user. The transparency of the resulting C4ISR architecture in a form intuitive to the Warfighter is the most important one of them. Using the easy to understand and intuitive products of the operational view, the influence of his various requirements, trade-offs in case of conflicts, the discovery of hidden causalities and unexpected interrelations become obvious. No distinction is needed between different domains of information technology, be it C4ISR or M&S. The composition of systems comprising the operationally necessary functionality delivered by components derived from the legacy C4ISR systems as well as M&S system – like already envisioned in [Tolk, 2000] – becomes possible, leading to better and more reliable systems.

VV&A in the context of heterogeneous systems can help to build a bridge between the participating and engaged communities. The use of a common architecture framework as a common basis is a necessary – hence not sufficient – first step.

## 4    Summary

The US DOD C4ISR Architecture Framework [US DOD, 1997] as well as his NATO counterpart [NATO, 2000] can contribute to facilitate the processes of VV&A in the domain of military M&S. Whether the products already comprised are sufficient or whether additional products in form of a V&V architectural view are necessary cannot be answered without additional research. The benefits, however, would manifest as improvements for the M&S community as well as for the C4ISR community and would doubtlessly increase the support for the Warfighter.